\documentclass[sigconf,nonacm]{acmart}

\setcopyright{none}
\usepackage{url}
\renewcommand{\normalsize}{\fontsize{10pt}{12pt}\selectfont}
\AtBeginDocument{%
  \setlength{\textheight}{8.7in}   
  \setlength{\topmargin}{0in}
  \setlength{\headheight}{12pt}   
  \setlength{\headsep}{12pt}      
  \setlength{\footskip}{24pt}     
}

\makeatletter
\def\@oddhead{}
\def\@evenhead{}
\def\ACM@authordraft{false}  
\makeatother

\usepackage{enumitem}
\usepackage{amsmath}
\usepackage{graphicx}
\usepackage{caption}

\AtBeginEnvironment{figure}{\vspace{-4pt}}
\AtBeginDocument{%
  }

\author{Wentao Hong}
\renewcommand{\shortauthors}{Wentao Hong}

\usepackage{siunitx}
\renewcommand\footnotetextcopyrightpermission[1]{}

\begin{document}
\title{AgentLens: Framework-Aware Characterization and Resource Control for Heterogeneous LLM Agents}



\begin{abstract}
LLM-based agents are increasingly deployed at scale, yet how they behave as host-side system workloads remains poorly understood. We present AgentLens, a cross-framework characterization spanning 10 frameworks and 11 matched coding tasks, instrumented across API boundaries, tool calls, process creation, memory composition, and microarchitectural events, together with a framework-aware coordination architecture for co-located agent instances. Our measurements show that framework orchestration structure, not task content, is the dominant source of resource variance. Framework identity explains 52.95\% of peak RSS variance while task content explains only 4.22\%, and MetaGPT triggers 117.2 \texttt{execve} events per tool call against 3.0 in CAMEL, a 39$\times$ gap on identical tasks. Resource dimensions are also systematically decoupled; the Spearman correlation between peak RSS and subprocess count is only 0.13, making no single metric a reliable proxy for the others. These properties render existing control approaches unsound for co-located agent workloads. Static allocation has no representation of the alternating API-wait and tool-execution phase structure that governs every run. Controllers lacking framework priors cannot distinguish whether observed pressure is structural or transient. Historical percentiles conflate stable framework fingerprints, task-driven variance, and within-run retry dynamics into a figure whose origin cannot be recovered. AgentLens addresses all three with offline framework fingerprints and a phase-aware resource controller that applies differentiated memory coordination through cgroup v2 without instrumenting framework internals. In trace-driven replay across three co-location pairs, AgentLens bounds the protected worker's P95 allocation latency to 5.7--7.1\,ms when aggregate demand exceeds the parent budget by 10\%, a $2.1\times$--$2.6\times$ improvement over static allocation.

\end{abstract}

\maketitle

\section{Introduction}
\label{Sec:Introduction}

Large language model(LLM) agents are increasingly used to solve open-ended tasks from software engineering\cite{jain2024livecodebench,badertdinov2025swerebench,swe-agent}, research\cite{openai2025deepresearch, google2025gemini}, web automation\cite{he2024webvoyager, zhou2023webarena} to interactive social tasks\cite{park2023generative, xie2024aimetropolis}.
Rather than serving a single inference, agents typically operate in a ReAct-style loop\cite{yao2023react, sarukkai2025selfgenerated}, alternating between reasoning about the next step and executing actions such as tool calls, environment interactions, and state updates in sandboxed containers\cite{dworken2025beyond, anthropic2026secure}. As agents are being rapidly deployed in production environments, with major technology providers offering them as managed services and co-locating large numbers of instances on shared hosts, efficient resource management has become a first-order systems concern\cite{github2026copilot, openai2025codex, chen2025kairos, laju2026nalar, bodea2025trusted, google2025jules, cognition2024devin}.

Existing research on agent systems has largely followed two directions. One focuses on inference-side efficiency, such as maximizing token throughput\cite{kwon2023vllm}, KV cache management, and GPU utilization. The other focuses on agent-serving architectures\cite{laju2026nalar, ni2025astraea, chen2025kairos}, such as multi-agent request scheduling, session management, state hosting, and elasticity. These efforts have substantially advanced agent-serving infrastructure. However, they typically treat the agent as an opaque containerized workload to be scheduled\cite{laju2026nalar, li2025continuum, ni2025astraea}, leaving the specific resource demands and execution patterns of the underlying agentic processes largely uncharacterized.

As a growing share of agent execution cost shifts from GPU inference to the host\cite{rajCPUCentricPerspectiveAgentic2025}, this gap becomes increasingly consequential. Yet without a characterization of how agent execution actually behaves on the host, it is impossible to determine whether existing control interfaces designed for request- or container-level workloads can manage agent resource behavior correctly. Recent work such as AgentCgroup\cite{zheng2026agentcgroup} has begun to analyze agent runtimes at the operating-system level and has proposed tool-boundary-aware resource control. However, its scope remains limited to a single framework and intra-agent resource management. A systematic characterization across multiple frameworks and resource dimensions, and a control mechanism that accounts for framework-specific execution structure, remains missing.

First, agent execution is request-paced and bursty instead of continuous. It alternates between intensive tool execution and remote LLM reasoning to produce sharp transitions in host-side demand. The temporal structure of these bursts varies across frameworks because orchestration logic is the dominant source of variance in burst shape and frequency. Second, tool execution creates divergent host-side pressure depending on how each framework orchestrates tasks. Identical agent semantics produce fundamentally different resource footprints. For example, MetaGPT spawns 39× more processes than CAMEL on the same task. These divergent realization paths translate the same logical action into different process structures and kernel overhead across all frameworks.
Third, agent workloads exhibit systematic and asymmetric decoupling across resource dimensions. RSS accumulates across rounds and remains elevated during LLM reasoning phases while CPU utilization is near zero. The Spearman correlation between CPU time and peak RSS is only 0.54 and drops to 0.13 between peak RSS and subprocess count. This confirms that single-metric profiling is an insufficient basis for characterizing or provisioning agent workloads.

First, static and equal-split allocation schemes fail because they assume stable resource demand over time. Agent execution instead follows alternating phases with fundamentally different resource profiles. Setting quotas to peak demand wastes resources during idle phases while setting to average causes systematic under-provisioning during bursts. Since these schemes cannot track the phase structure of execution, the mismatch between capacity and demand becomes a structural property of every agent run. Second, existing control approaches lack visibility into framework-specific execution structures. This makes proactive differentiated provisioning impossible regardless of control granularity. The absence of such visibility forces all control decisions to be reactive. Systems respond uniformly to aggregate pressure only after degradation occurs instead of pre-positioning resources according to the specific demand structure of each framework. Third, predictive approaches incorrectly assume that historical consumption reliably forecasts future demand. Agent consumption is jointly shaped by framework orchestration style, task characteristics, and runtime variability. Existing approaches treat run-level consumption as a unified signal without distinguishing stable structural properties from variable execution outcomes. Consequently, forecasts degrade precisely when framework and task diversity is highest.

We present AgentLens, a framework-aware, phase-aware resource coordination framework for co-located AI coding agents. AgentLens builds on a cross-framework host-side characterization that aligns semantic execution events with host-visible telemetry such as cgroup counters\cite{heo2015cgroupv2}, subprocess activity\cite{gregg2020bpf}, PSI signals\cite{weiner2018psi}, and microarchitectural coun-
ters\cite{weaver2013linux}, to recover how different frameworks translate similar agent semantics into different resource behaviors, and uses the resulting evidence to derive offline framework fingerprints, runtime phase signals, and a coordinated control layer for memory, CPU, and PID resources. 

This paper makes the following contributions:
\begin{itemize}
    \item First, we conducted cross-framework characterization of host-side resource behavior across 10 AI coding agent frameworks along three axes: execution temporal structure, tool-call-induced overhead, and resource usage patterns.
    \item Second, we identified three fundamental mismatches between existing resource-control abstractions and agent workload structure, arising in the temporal, structural, and predictive dimensions of control.
    \item Third, We proposed a framework- and phase-aware resource controller pairing offline fingerprints. Trace-driven replay shows $2.1\times$--$2.6\times$ lower protected P95 latency under 10\% over-subscription.
\end{itemize}

\section{Background and Motivation}                       
\label{Sec:Background}   

\subsection{Common Execution Model for Agent}              
Most modern agent frameworks follow a broadly similar ReAct-style\cite{yao2023react} execution pattern. In each round, the framework observes the current environment state, reasons about the next action, and executes that action locally or delegates it to an external tool. The result is fed back into the next round, forming a closed loop that continues until the task is complete or a termination condition is met. From the host perspective, this loop produces a recurring pattern of idle waiting during remote inference and concentrated bursts of local activity when tool calls, state updates, and orchestration logic execute on the host. While the high-level loop is shared across frameworks, they differ substantially in how they materialize these phases through tool dispatch logic, state retention depth, and the balance of work between remote inference and local execution.

Coding agents impose substantial local pressure through repetitive environment inspection and tool invocation. The framework resolves tool calls generated by the model through a registry and dispatches them as local operations, realized as shell commands, Python interpreters, or adapters specific to the framework. These operations execute within sandboxed environments and frequently materialize as subprocess trees involving process forking, environment setup, and filesystem interaction. The same logical action can therefore produce very different host-side footprints depending on how a framework implements its tool dispatch and execution path.

As agent deployments move into production, providers increasingly co-locate multiple agent instances on shared hosts to improve utilization~\cite{github2026copilot, openai2025codex, chen2025kairos}. Under co-location, the host must simultaneously manage instances from heterogeneous frameworks, each with its own realization path, state retention policy, and burst profile. This makes the resource management problem strictly harder than managing a single homogeneous workload. Pressure from structurally different sources arrives concurrently, and a control decision that is appropriate for one framework may be harmful for another sharing the same budget.
\subsection{Why Agents Defy Resource Models}                                                  
Traditional resource abstractions are built around execution units that are naturally bounded by a request, a task, or a stable process structure~\cite{barroso2003web, dean2008mapreduce, jonas2019cloud}. In those settings, allocation, attribution, and isolation can be attached to a well-defined object whose resource footprint is either short-lived or structurally stable. Agents do not expose such a unit. End-to-end progress is driven by repeated remote model interaction, but the host-side cost that matters for control is released intermittently through tool execution, environment inspection, and framework-level orchestration. Execution state persists across rounds rather than remaining confined to a single request, and the volume and structure of local work varies substantially depending on model output and framework implementation. The object that the host must manage is therefore neither a request-bounded inference client nor a structurally stable long-lived process.

What the host observes instead is a sequence of local effects, including environment inspection, state mutation, iterative verification, and the accumulation of runtime state across iterations. Focusing on inference alone obscures the host-side work performed between model calls, while a conta-
iner-level view fails to capture how pressure originates from tool execution, state updates, and framework-specific realization paths. Neither abstraction exposes the execution structure that resource control requires. Under co-location, this gap becomes more consequential. Without visibility into how each framework realizes its execution, a shared controller cannot distinguish whose pressure is causing contention or respond in a framework-appropriate way.

\vspace{-1em}
\subsection{The Host-Side Observability Gap}

This gap is compounded by how agent frameworks are currently evaluated. Prior work focuses primarily on benchmark pass rate, task success, end-to-end runtime, and token cost\cite{jain2024livecodebench, badertdinov2025swerebench, mialon2024gaia, liu2023agentbench, 2023opencompass}. These metrics are necessary for understanding utility but do not expose the host-side execution behavior on which resource control decisions depend. They cannot reveal when host-side pressure appears, what causes it, or how it differs across frameworks running identical tasks.

At the host level, the relevant behavior is physical rather than semantic. What matters is the timing of API-wait intervals versus local execution bursts, the subprocess activity during tool dispatch, and the accumulation of memory across rounds. Without a systematic account of these dynamics, it is impossible to determine whether a framework is inherently externally paced, locally bursty, or coordination-heavy, or to understand how framework choice shapes the resource management problem under co-location. This is the gap that our characterization addresses.

\section{Measurement Methodology}
\label{sec:methodology}

\subsection{Framework and Task Selection}
We study 10 agent frameworks: Agno\cite{agno_2024}, AutoGen\cite{autogen}, CAM-
EL\cite{camel_2023}, CrewAI\cite{crewai_2024}, LangGraph\cite{langgraph_2024}, MetaGPT\cite{hong2024metagpt}, 
Swarm\cite{openai_swarm_2024}, PydanticAI\cite{pydantic_ai_2024}, Smolagents\cite{smolagents_2024}, and SWE-agent\cite{swe-agent}. The set includes lightweight single-agent orchestrators with minimal runtime overhead, tool- and process-heavy coding agents that materialize actions through subprocess chains, orchestra-
tion- and memory-heavy runtimes that retain substantial context state across rounds, and explicit multi-agent frameworks with staged coordination structure.

To control task semantics, we evaluate all frameworks on a matched set of 11 SWE-rebench tasks selected to span a range of repository sizes, patch complexity, and required tool interaction depth. The matched-task design holds the application domain and problem family sufficiently constant so that cross-framework host-side differences remain attributable to framework orchestration rather than task variation. 

\subsection{Experimental Setup}

All experiments are conducted on a dedicated Linux server running kernel 6.12.28 with cgroup v2 and PSI enabled. The host platform is a 4-socket Intel Xeon Gold 6448H system with 256 logical CPUs and 1.5 TiB of RAM. Each framework run executes inside an isolated container so that process, memory, PID, and pressure signals can be attributed through a stable cgroup boundary. This setup mirrors the sandboxed execution model commonly used by modern coding agents while also providing a consistent attribution unit for host-side measurement. Model inference is issued through the remote DeepSeek API\cite{deepseek_api_2025} so that the host bears only the cost of orchestration, tool execution, subprocess creation, and context-state management. Remote API latency introduces variance in wall-clock run duration, but the structural metrics we analyze are determined by framework orchestration logic rather than by the absolute timing of API responses. 

\subsection{Host-Side Instrumentation and Derived Metrics}

AgentLens follows a top-down cross-layer measurement methodology. The semantic layer exposes what the framework intends to do but not what the host actually pays. Cgroup time series capture aggregate resource consumption but miss short-lived subprocess activity that falls between sampling intervals. Kernel-event tracing fills this gap but does not expose microarchitectural cost. And microarchitectural counters quantify execution efficiency but carry no semantic context. Each layer therefore addresses a blind spot left by the others, and the findings in this paper depend on correlating signals across all four.

At the semantic layer, we record lifecycle markers, API send/receive events, tool events, and explicit multi-agent events such as \texttt{AGENT\_HANDOFF}. These traces expose when a runtime is waiting on the remote model API, when it returns to the host for local work, and how coordination events are interleaved with tool execution. At the container boundary, we collect cgroup time series over memory, CPU, I/O, PID count, and PSI. At the kernel-event layer, we use \texttt{bpftrace} to capture short-lived subprocess activity that falls between cgroup sampling intervals and would otherwise be invisible to coarse-grained collection. At the microarchitectural layer, we use \texttt{perf stat} to gather counters such as instructions, cycles, cache misses, page faults, context switches, and CPU migrations. All signals are synchronized to a common monotonic time base. This unified timeline allows us to precisely align high-level semantic phases with low-level system footprints.

From these traces, we derive metrics along three complementary dimensions that together characterize when host-side pressure appears, what form it takes, and how it is released. Temporal metrics include API-wait ratio and the timing and duration of local execution bursts. Resource metrics include CPU consumption, \texttt{cpu\_avg\_cores}, RSS and anonymous memory footprint, PID usage, and \texttt{execve} density. Pressure metrics include PSI and throttling signals that capture when resource contention becomes visible at the runtime boundary. Together these dimensions allow us to compare frameworks not only by aggregate resource usage but also by the structure and timing of the host-side cost they impose.

\section{Characterizing Agent Workloads}

\subsection{Temporal Structure}

\label{Sec:Temporal Dynamics}

\subsubsection{Runtime Phase Breakdown Across Frameworks}


To characterize where end-to-end runtime is spent, we decompose each run into three mutually exclusive phases derived from API-boundary and tool-call instrumentation events. A run is in \texttt{api\_wait} while a model request has been issued and the response has not yet returned. It enters \texttt{tool\_exec} once a tool-call interval begins and exits when the interval ends. Within \texttt{tool\_exec}, we isolate a \texttt{retry\_spiral} sub-component comprising time from tool-call intervals that return non-zero exit codes and from the start of the most recent tool invocation to any \texttt{API\_CALL\_ERROR} event in the same round, bounded above by total \texttt{tool\_exec} time; the figure breaks this sub-component out as a separate bar adjacent to \texttt{tool\_exec} to make failure-driven overhead visible. The \texttt{planning} phase is the residual computed as total run duration minus \texttt{tool\_exec} and \texttt{api\_wait}, capturing response parsing, prompt assembly, and inter-round state updates.

\begin{figure}[t!]
    \centering
    \includegraphics[width=\linewidth]{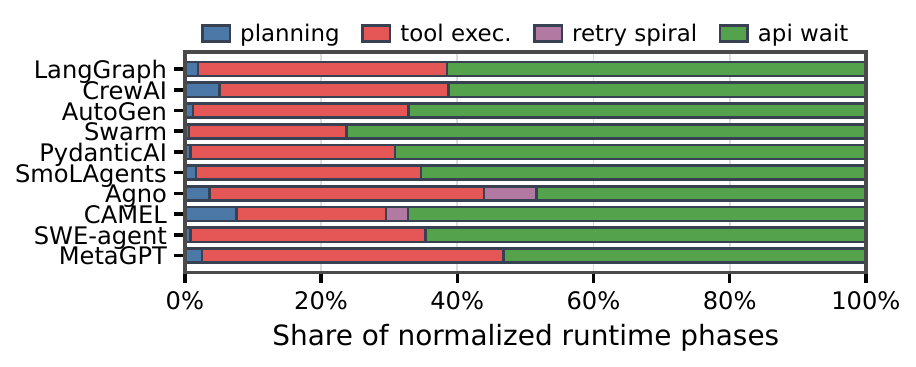}
    \Description{Horizontal stacked bars compare the runtime fractions spent planning, executing tools, retrying, and waiting for APIs across ten agent frameworks. API waiting occupies the largest share.}
    \caption{Normalized runtime phase composition across agent frameworks, broken down into planning, tool execution, API waiting, and retry activity.}
    \label{fig:api_temporal_master}
    \vskip -5pt
\end{figure}

Figure~\ref{fig:api_temporal_master} shows that API waiting remains the largest phase across all frameworks, but tool execution represents a non-trivial and variable fraction of total runtime. The API waiting share ranges from approximately 47\% in Agno to 78\% in Swarm, while tool execution accounts for roughly 22\% to 43\%, with MetaGPT and Agno at the high end. Planning occupies a small but visible share in CAMEL and CrewAI and is negligible in most other frameworks. Retry spiral activity appears only in Agno and CAMEL. The dominant temporal envelope is remote inference waiting, but host-side tool execution is far from negligible and varies substantially across frameworks.

\subsubsection{A Request-Paced and Bursty Execution Rhythm}

\begin{figure}[t!]
    \centering
    \includegraphics[width=\linewidth]{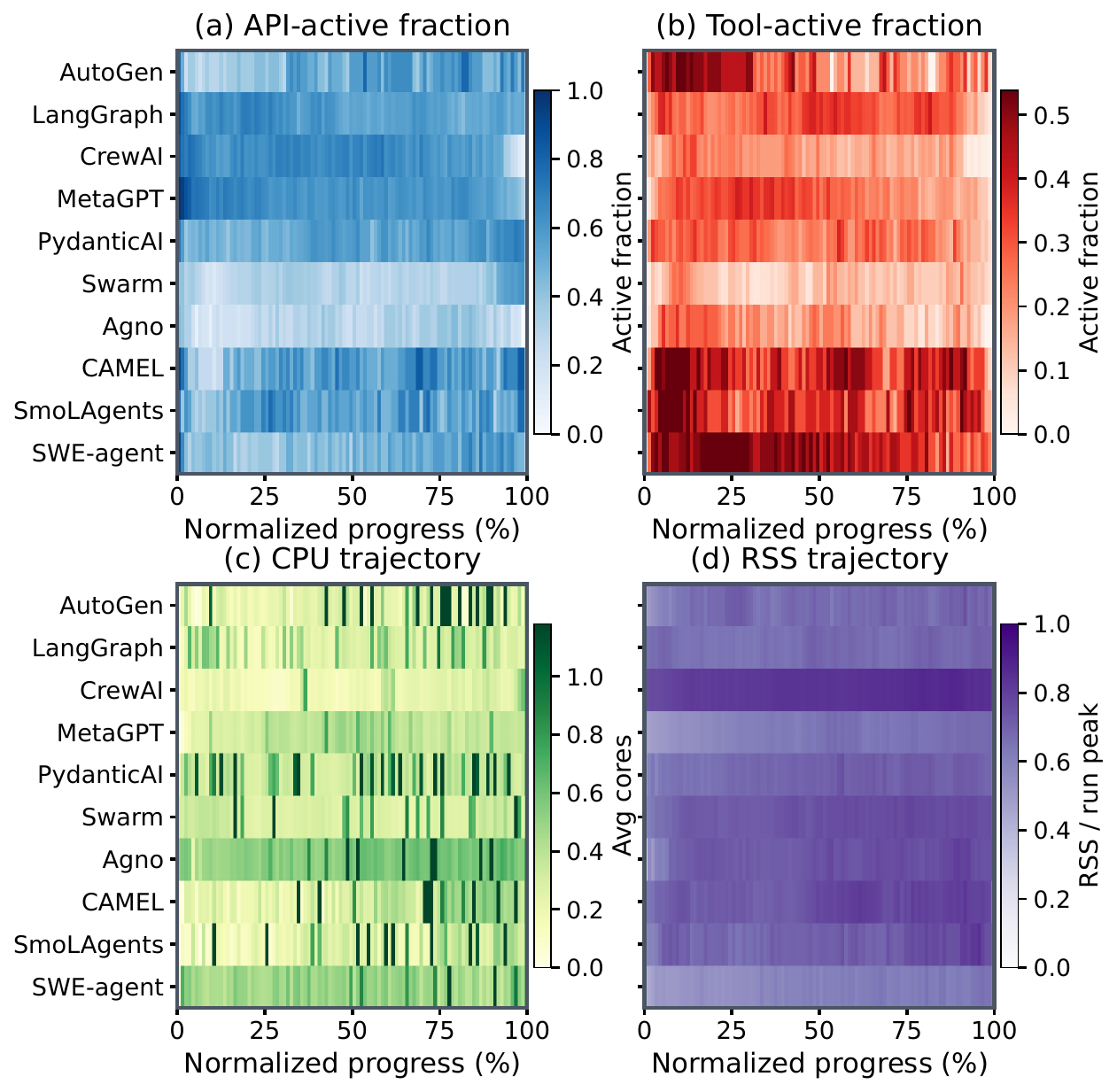}
    \Description{Four heatmaps align ten frameworks along normalized run progress. The panels show API activity, tool activity, CPU usage, and RSS relative to each run peak.}
    \caption{Aligned time-series traces of API activity, tool-execution intervals, CPU utilization, and RSS over representative runs across agent frameworks.}
    \label{fig:timeseries-trajectory}
    \vskip -5pt
\end{figure}

To move beyond aggregate API-time ratios, we align cgroup time-series samples with API and tool-call timestamps and then normalize each run onto a common $[0,1]$ progress axis. Figure~\ref{fig:timeseries-trajectory} aligns API activity, tool-active intervals, CPU usage, and RSS over a normalized run-progress axis for representative executions from each framework. Panels (a) and (b) show a consistent temporal pattern across frameworks. Execution is request-paced across frameworks, with API-active phases alternating with phases of host-side tool execution. MetaGPT, SWE-agent, CAMEL, and AutoGen show the most pronounced temporal concentration, but in different forms. SWE-agent and AutoGen are most front-loaded, CAMEL shows both early and late concentration, and MetaGPT sustains heavier local activity over a larger fraction of the run. Agno and Swarm, by contrast, distribute tool activity more evenly and at lower density, while the remaining frameworks fall between these two extremes.

These differences are consistent with the execution structures of the frameworks. SWE-agent front-loads host-side work because its observation-driven action loop concentrates repository inspection, command execution, and hypothesis checking early in the run. AutoGen also tends to front-load tool activity, as tool calls are often emitted in early phases and executed concurrently when returned together. CAMEL exhibits a more stage-structured trajectory, with early decomposition and setup and later synthesis and verification, producing concentration near both ends of the run. MetaGPT sustains heavier local activity over a broader portion of execution because planning, message passing, and task propagation are repeatedly interleaved across roles. By contrast, Agno and Swarm rely on lighter-weight execution loops, so tool invocations are distributed more evenly and at lower density across run progress. The remaining frameworks largely occupy intermediate positions between these extremes, preserving the same request-paced burst structure but differing less sharply in where local activity concentrates over execution.

Panels (c) and (d) extend the API and tool-activity view with host-side resource trajectories. CPU demand appears in short, concentrated bursts that closely track periods of local execution rather than remaining sustained across the full run. RSS evolves on a different timescale, changing more gradually and often remaining elevated between execution rounds. These two trajectories reflect distinct aspects of the host-side  footprint. One captures transient bursts driven by tool execution, and the other captures persistent state accumulated between those bursts. Frameworks differ substantially in both dimensions. MetaGPT maintains a higher RSS baseline throughout execution, consistent with its role-based multi-agent architecture that retains state across multiple interleaved agents. Agno, by contrast, shows similar CPU burst patterns but a substantially flatter RSS profile, consistent with lighter state retention between invocations.


\subsection{Tool Execution Overheads across Frameworks}
\label{Sec:Tool Execution}

\subsubsection{Command Composition and Host-Side Cost}
\label{sec:tool_call_analysis}

\begin{figure}[t!]
    \centering
    \includegraphics[width=1.0\linewidth]{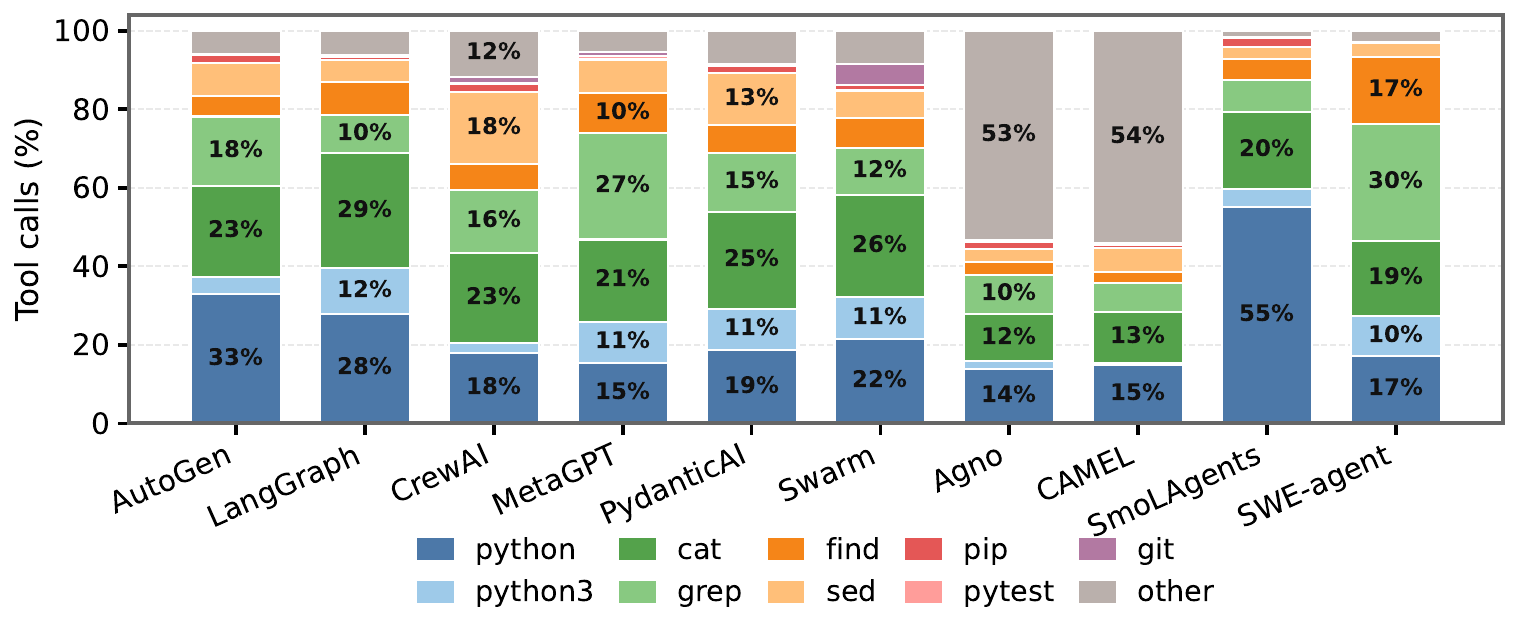}
    \Description{Stacked bars compare command composition across ten frameworks, distinguishing Python, file inspection, package installation, version control, testing, and other commands.}
    \caption{Command composition of realized tool calls across agent frameworks.}
    \label{fig:tool-call-granularity_a}
\end{figure}

\begin{figure}[t!]
    \centering
    \includegraphics[width=0.99\linewidth]{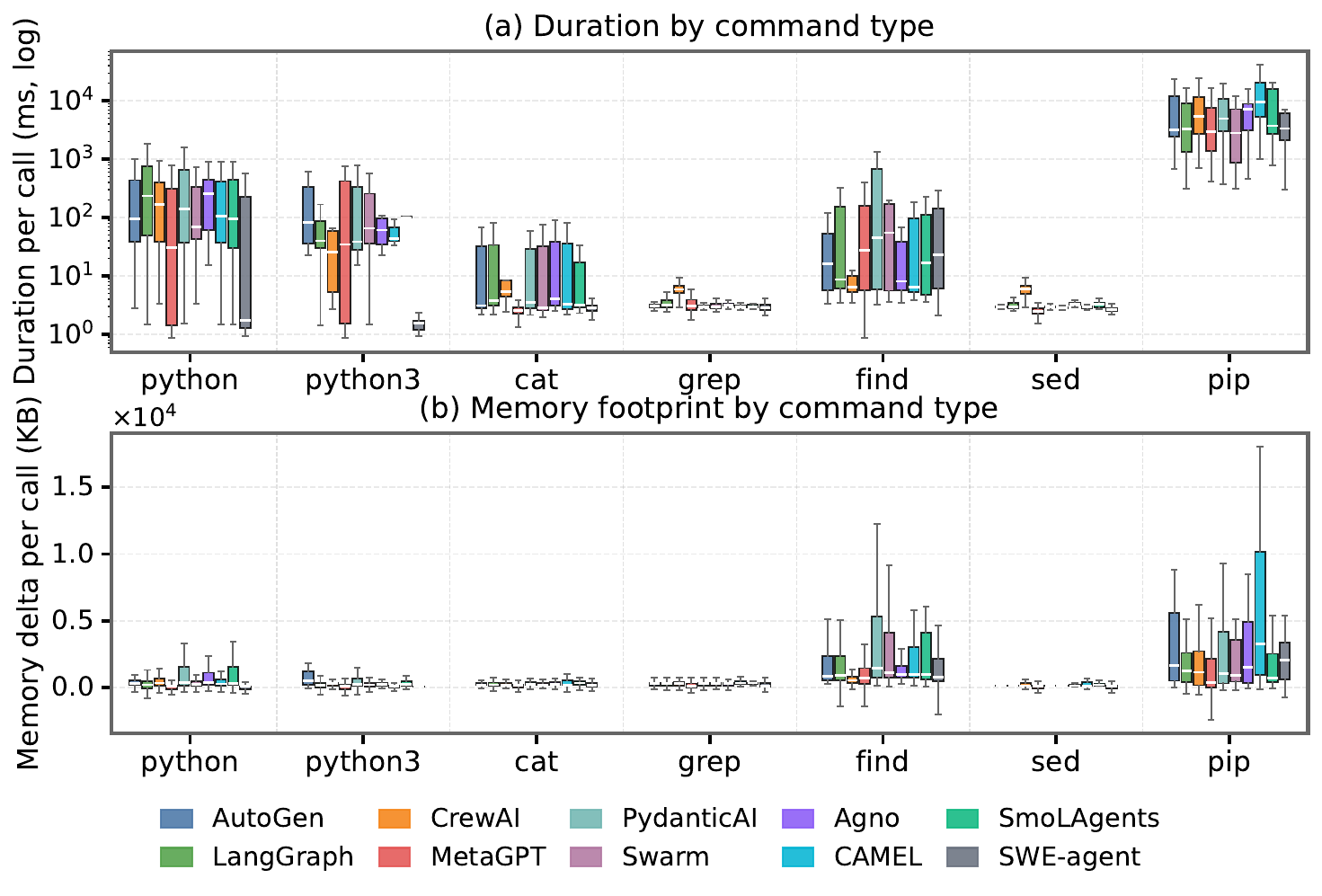}
    \Description{Two panels compare command types across frameworks. The first shows per-call duration on a logarithmic scale; the second shows per-call memory change. Python and pip calls have higher costs than file inspection commands.}
    \caption{Per-command host-side cost within tool-call windows across agent frameworks.}
    \label{fig:tool-call-granularity_bc}
    \vskip -5pt
\end{figure}

Since host-side resource bursts concentrate within tool-call windows, run-level statistics cannot reveal the composition of local cost. All frameworks support tool use but differ in how they define and realize a single tool call on the host, so we take each framework's own tool call as the primary unit of analysis. We classify each call by the most common commands executed within that window.  

Figure~\ref{fig:tool-call-granularity_a} shows that command composition varies substantially across frameworks. In SmoLAgents, \texttt{python} accounts for 55\% of tool calls, reflecting an interpreter-centric execution path. In SWE-agent, \texttt{grep} makes up 30\%, \texttt{cat} 19\%, and \texttt{find} 17\%, reflecting a tool set oriented toward systematic repository navigation. In AutoGen, LangGraph, and MetaGPT, \texttt{python} accounts for 26 to 40\% while \texttt{grep}, \texttt{cat}, and \texttt{find} together contribute 40 to 58\%, reflecting tool sets that mix interpreter invocations with direct file inspection without either dominating. In Agno and CAMEL, the \texttt{other} category represents 53\% and 54\% of calls respectively, reflecting tool calls realized through framework wrappers rather than direct shell commands.

The command distribution of agent frameworks is determined by what tools it exposes to the model and how those tools are implemented at the host level. A framework that routes work through a Python interpreter will show a distribution dominated by interpreter calls, while one that makes repository inspection explicit will naturally produce high shares of \texttt{grep}, \texttt{cat}, and \texttt{find}. The same logical action can therefore appear as a direct \texttt{python} or \texttt{grep} invocation in one framework and as a wrapper call with no recognizable shell command in another.

Command composition matters because different commands impose markedly different costs on the host. Figure~\ref{fig:tool-call-granularity_bc}(a) shows that execution-oriented commands such as \texttt{pip} and \texttt{python} are orders of magnitude slower than inspection primitives. Figure~\ref{fig:tool-call-granularity_bc}(b) shows that memory deltas follow the same ordering. The median \texttt{pip} call lasts 4212.1\,ms, compar-
ed with only 2.9\,ms for median \texttt{grep}, a gap of roughly three orders of magnitude. Even \texttt{python}, which is far more common than \texttt{pip}, remains about 14$\times$ slower than \texttt{grep} at the median, with 40.7\,ms versus 2.9\,ms. Memory deltas follow the same rank. Median \texttt{pip} calls in CAMEL reach about 3.2\,MB, whereas median \texttt{grep} calls in MetaGPT remain at 128\,KB.

Heterogeneity also persists within the same command. Two frameworks may both issue \texttt{python}-dominated calls yet exhibit markedly different realized cost. As shown in Figure~\ref{fig:tool-call-granularity_bc}(a), median \texttt{python} calls last about 30\,ms in MetaGPT but about 333\,ms in LangGraph. Figure~\ref{fig:tool-call-granularity_bc}(b) shows that median \texttt{python} memory deltas range from roughly 0.1\,MB in MetaGPT and SWE-agent to about 0.4\,MB in CrewAI. Command identity alone is therefore not an adequate abstraction for local host cost. The realized cost also depends on the framework-side execution path that surrounds the command.

\subsubsection{CPU Intensity within Tool-Call Windows}

\begin{figure}[t!]
    \centering
    \includegraphics[width=1.0\linewidth]{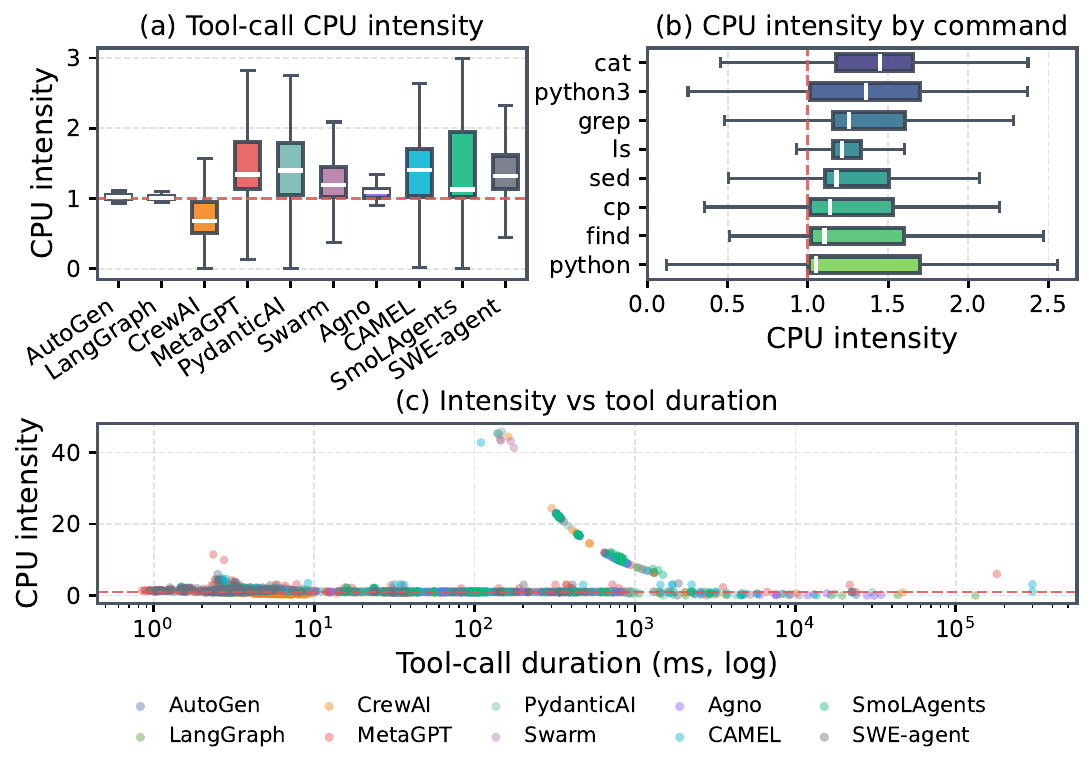}
    \Description{Three panels show tool-call CPU intensity by framework, CPU intensity by command, and the relationship between CPU intensity and tool-call duration. A reference line marks intensity equal to one.}
    \caption{CPU utilization measured during tool-call windows across agent frameworks.}
    \label{fig:tool_cpu_intensity}
    \vskip -5pt
\end{figure}

To quantify how much computation is packed into a tool call window, we define \emph{CPU intensity} as CPU time divided by wall-clock duration for that window. A value near 1.0 indicates that one CPU core is busy for most of the interval, values above 1.0 indicate parallel work across multiple cores or helper processes, and values below 1.0 indicate more waiting within the same boundary.

Figure~\ref{fig:tool_cpu_intensity} shows that tool-call windows are typically CPU-dense rather than lightweight background activity. Most frameworks cluster near or above the 1.0 boundary, meaning that a single tool call window often keeps one core busy for most of its duration. The upper end is led by CAMEL and PydanticAI, with median intensities of 1.41 and 1.40, consistent with local execution paths that pack more interpreter work, helper process activity, or wrapper side computation into each window. MetaGPT and SWE-agent follow at 1.34 and 1.33, indicating similarly dense local realization once tool execution begins. By contrast, AutoGen and LangGraph remain just above the same boundary at 1.02 and 1.01, which suggests that their tool call windows remain CPU-active but are less densely packed than the heavier local paths above. CrewAI remains well below the boundary at 0.68, consistent with more waiting or lighter-weight control work within the same interval.

CPU-dense execution is the norm once tool execution starts, not an exceptional case. This holds not only in median intensity but also in prevalence. Tool calls with intensity above 1.0 account for nearly all calls in PydanticAI, Swarm, SWE-agent, and MetaGPT, and the same remains true for most calls in AutoGen, LangGraph, Agno, CAMEL, and SmoLAgents, with CrewAI as the clear exception. The systems implication is that agent resource pressure has both a temporal and an intensity dimension. Workloads are temporally sparse, but the windows where execution occurs are computationally dense, and frameworks differ materially in how much CPU pressure they impose within each window.               
\begin{figure}[t!]
    \centering
    \includegraphics[width=1.0\linewidth]{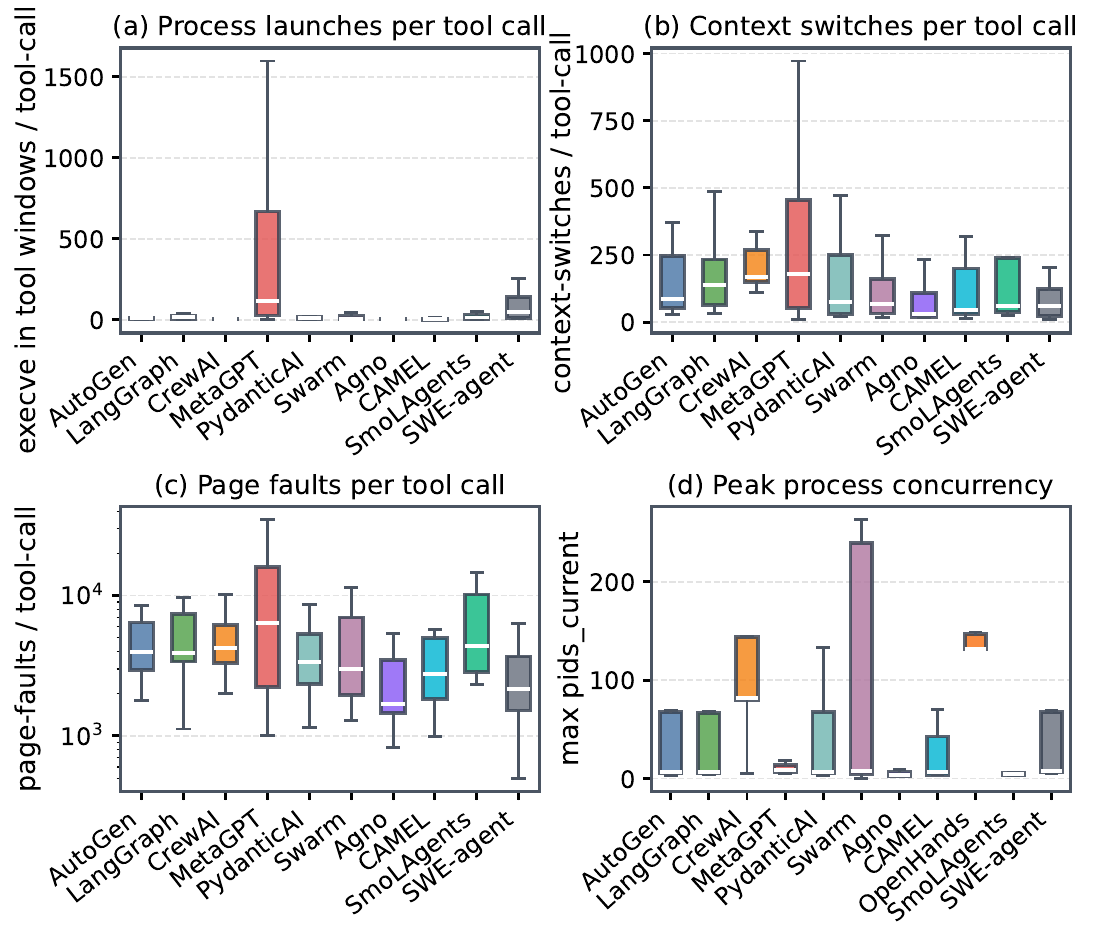}
    \Description{Four panels compare process launches, context switches, and page faults per tool call, followed by peak concurrent process counts, across agent frameworks.}
    \caption{Process-realization churn induced by tool execution across agent frameworks.}
    \label{fig:process_churn}
    \vskip -10pt
\end{figure}

\subsubsection{Tool Call Realization and Kernel Overhead}

While high tool-call intensity creates significant resource pressure, the actual OS impact depends on how frameworks realize tool calls at the host level. Frameworks differ in whether they absorb tool execution in-process or materialize it through repeated subprocess creation. Each subprocess event carries kernel-facing cost through \texttt{execve} calls, address-space setup, and scheduler pressure, and frameworks that rely on subprocess-based realization accumulate these costs at every tool invocation.

Figure~\ref{fig:process_churn} reveals wide variation in per-tool-call kernel overhead across frameworks. MetaGPT triggers a median of 117.2 \texttt{execve} events per tool call, compared with 51.6 in SWE-agent, 15.2 in SmoLAgents, 10.7 in AutoGen, and only 3.0 in CAMEL, a nearly 39x gap across frameworks performing the same abstract tool-use operation. MetaGPT also incurs a median of 6420 page faults and 180.6 context switches per tool call, indicating repeated address-space construction and scheduler intervention at each invocation. Figure~\ref{fig:process_churn}(d) reveals a separate pattern in peak process concurrency. Swarm reaches a median peak of approximately 230 concurrent processes per run and OpenHands approximately 150, far above all other frameworks, indicating that these frameworks involve substantially more simultaneous process activity than serial-spawning frameworks.

\begin{figure}[t!]
    \centering
    \includegraphics[width=0.99\linewidth]{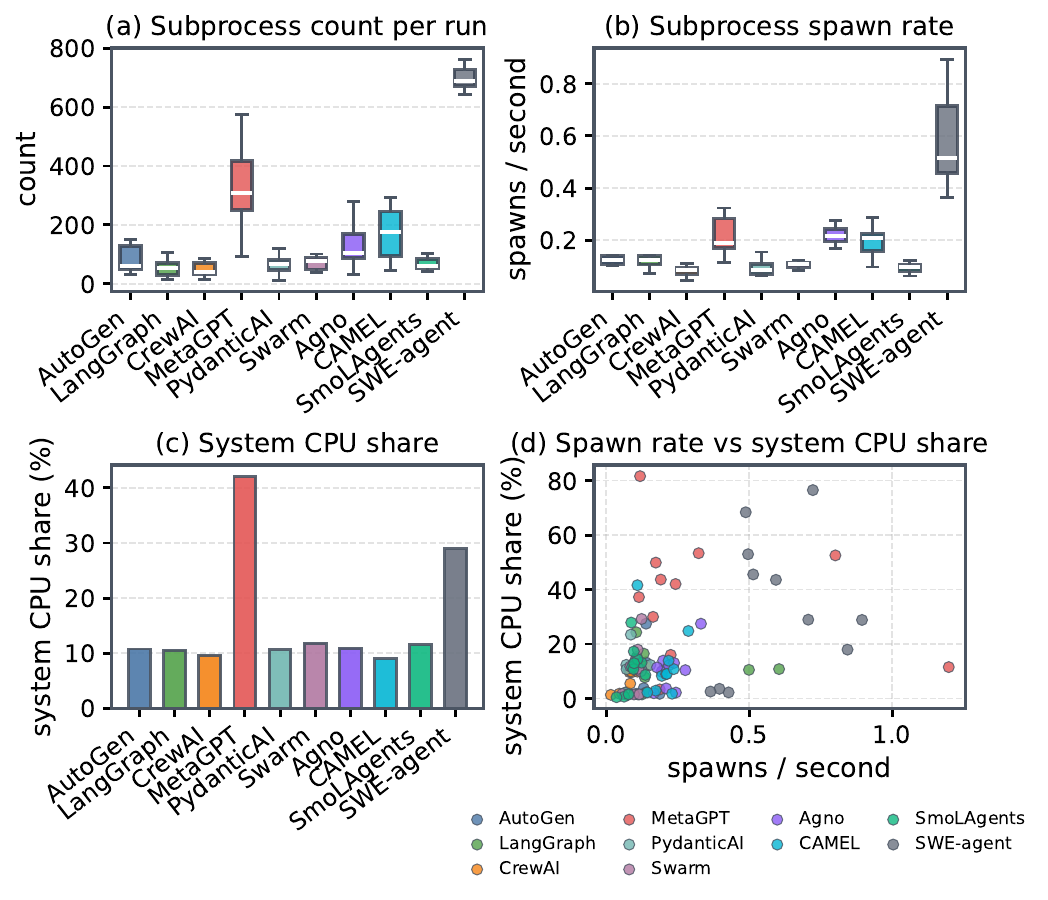}
    \Description{Four panels compare subprocess counts per run, spawn rates, and system CPU shares across frameworks, then plot spawn rate against system CPU share.}
    \caption{Kernel-level execution-overhead metrics collected during subprocess-intensive tool execution across agent frameworks.}
    \label{fig:subprocess-overhead}
    \vskip -10pt
\end{figure}

High spawn count does not imply high per-spawn kernel cost, and Figure~\ref{fig:subprocess-overhead} shows this separation clearly. SWE-agent accumulates the most subprocesses per run at a median of  724.5 and the highest spawn rate at 0.5 spawns per second, with MetaGPT ranking second at approximately 308.5 subprocesses per run and 0.22 spawns per second.Yet MetaGPT spends approximately 42\% of wall-clock time in system mode and 76.2\% of its CPU time in kernel mode, compared with 30\% and 47.9\% respectively for SWE-agent, despite SWE-agent spawning at more than twice the rate. Figure~\ref{fig:subprocess-overhead}(d) confirms that spawn rate and system CPU share are not linearly related. MetaGPT concentrates high system-mode CPU pressure at moderate spawn rates, while SWE-agent sustains high spawn rates at lower per-spawn kernel cost. These two frameworks therefore represent distinct subprocess overhead profiles. One is driven by spawn volume with moderate per-spawn cost, and the other by per-spawn weight at moderate spawn volume. Treating subprocess count alone as a proxy for kernel overhead would mischaracterize the cost structure of both.

Tool execution is a framework-governed activity in which command composition, per-invocation computation density, and subprocess realization cost all vary independently. A single run-level metric cannot simultaneously capture all three dimensions, and the dimensions do not collapse into a common ordering across frameworks. Seemingly identical agent workloads can therefore impose radically different host-side pressure depending solely on which framework realizes them.

\subsubsection{Tool Call Failures across Frameworks}
\label{Sec:Failure and Recovery}

When a tool call fails, the runtime often pays again for environment probing, wrapper entry, package installation, repository inspection, and verification retries even though the task itself has not advanced. 



\begin{figure}[t!]
    \centering
    \includegraphics[width=\linewidth]{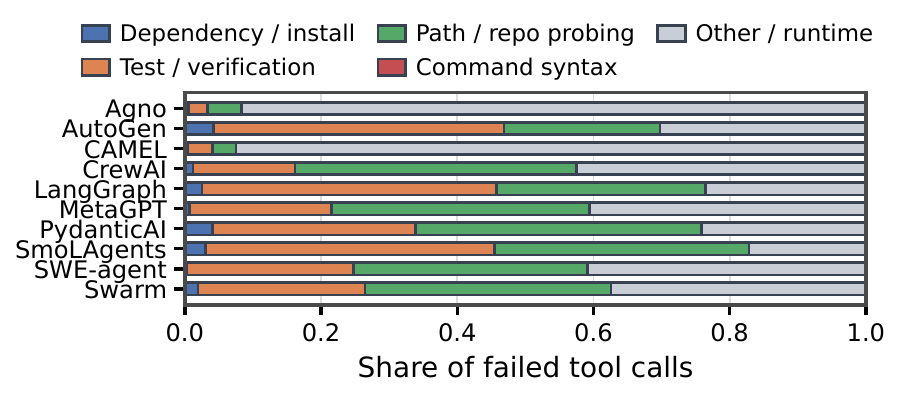}
    \Description{Horizontal stacked bars show the fractions of failed calls attributed to dependency installation, testing, path probing, command syntax, and other runtime failures for each framework.}
    \caption{Failure type composition of failed tool calls across agent frameworks.}
    \label{fig:supp-failure-path-categories}
    \vskip -5pt
\end{figure}

\begin{figure}[t!]
    \centering
    \includegraphics[width=\linewidth]{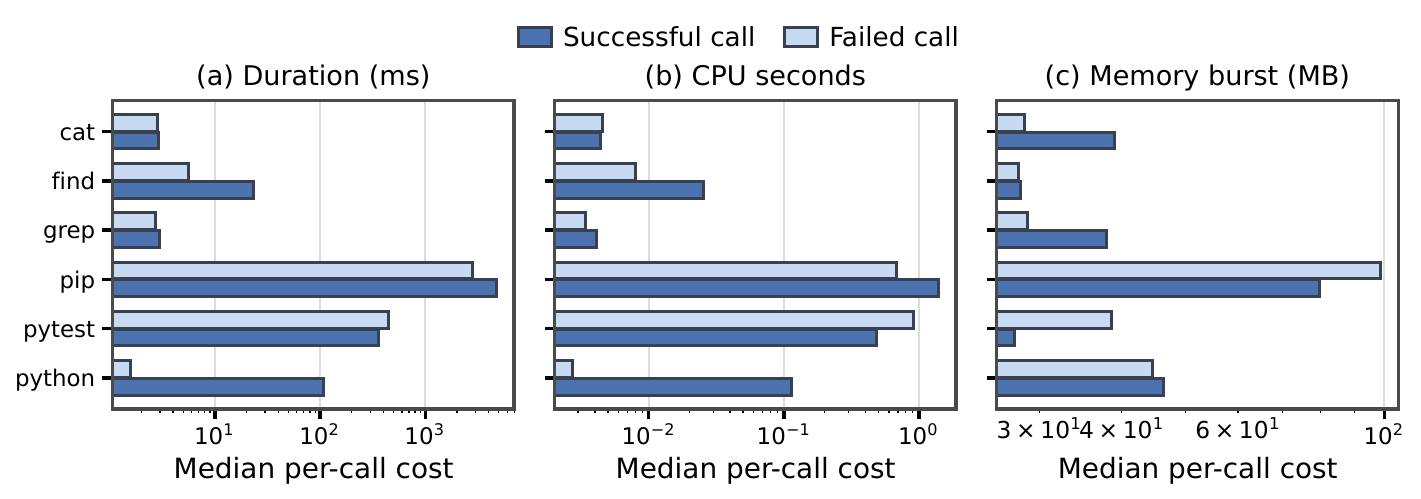}
    \Description{Three panels compare successful and failed cat, find, grep, pip, pytest, and python calls by median duration, CPU seconds, and memory burst.}
     \caption{Per-call resource cost conditioned on outcome for each command type:(a) duration, (b) CPU seconds, and (c) memory burst, comparing successful and failed calls at the median.} \label{fig:supp-failure-conditioned-cost}
    \vskip -5pt
\end{figure}

\begin{figure*}[t!]
    \centering
    \begin{minipage}[t!]{0.64\textwidth}
        \centering
        \includegraphics[width=\textwidth]{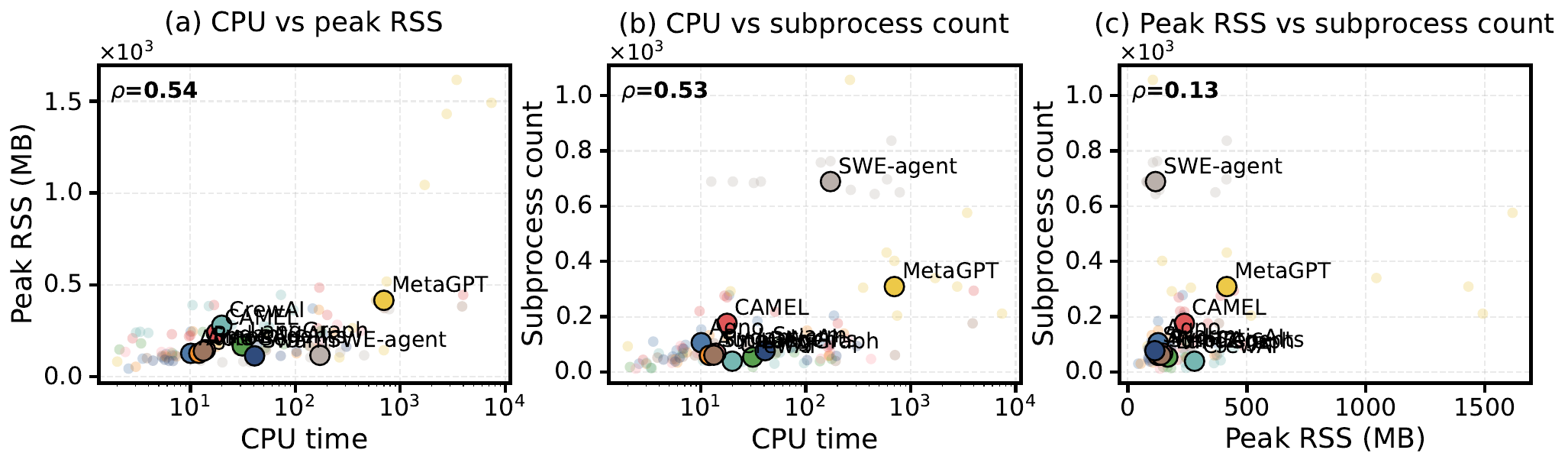}
    \Description{Three scatter plots compare CPU time with peak RSS, CPU time with subprocess count, and peak RSS with subprocess count. Points are grouped by framework; displayed correlations are 0.54, 0.53, and 0.13.}
        \captionof{figure}{ Correlations between CPU core-seconds, peak RSS, and subprocess count across agent frameworks.}
        \label{fig:local-metric-pairs}
    \end{minipage}
    \hfill
    \begin{minipage}[t!]{0.34\textwidth}
        \centering
        \includegraphics[width=\textwidth]{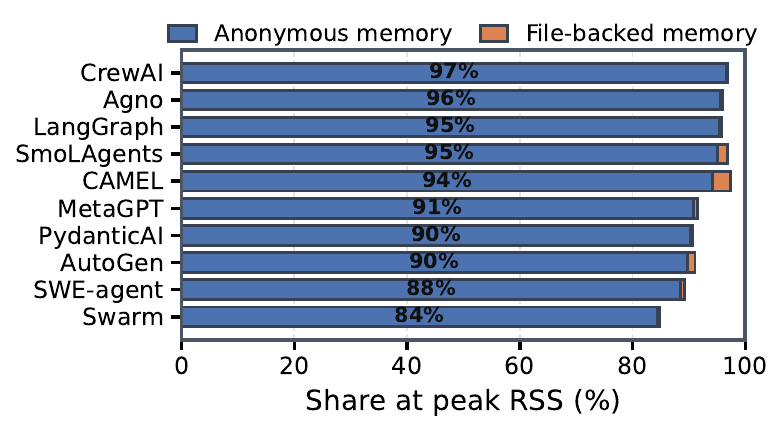}
    \Description{Horizontal stacked bars divide peak RSS into anonymous and file-backed memory for ten frameworks. Anonymous memory accounts for 84 to 97 percent.}
        \captionof{figure}{ Composition of peak RSS across agent frameworks.}
        \label{fig:memory-composition}
    \end{minipage}
\end{figure*}
Our evaluation reveals that this recovery burden varies drastically across frameworks. While most frameworks (e.g., Agno, PydanticAI, LangGraph) maintain median tool success rates above 87\% with fewer than 10 errors per run, MetaGPT and SWE-agent emerge as extreme outliers. These two frameworks exhibit median error counts of 118 and 181.5 per run, respectively. Crucially, their recovery paths are highly resource-intensive, as each failure triggers a heavy sequence of missing-package installation, test harness discovery, and command syntax repair—effectively re-entering the same costly tool-realization paths as successful executions.

Figure~\ref{fig:supp-failure-path-categories} shows that these failures are not
homogeneous. Some frameworks spend a larger share of failed activity on
dependency repair and installation, some on verification and test reruns, and
others on repository or path probing. This split again reflects implementation
choices. Frameworks that expose shell-heavy repair paths tend to revisit
installation and probing loops, whereas frameworks with stronger verification
stages more often pay for repeated test execution. The host-side recovery
burden therefore depends not only on how often a framework fails, but also on
which recovery path it repeatedly enters.

Figure~\ref{fig:supp-failure-conditioned-cost} shows that the per-call resource cost of a failed invocation depends heavily on command type. For heavy commands such as \texttt{pip} and \texttt{pytest}, failed calls consume nearly as much resource as successful ones. Median failed \texttt{pip} calls last approximately 3000\,ms and consume approximately 0.8 CPU seconds, compared with 4000\,ms and 1.0 CPU seconds for successful calls, and failed \texttt{pip} calls produce a larger memory burst than successful ones. Failed \texttt{pytest} calls consume comparable CPU to successful calls. By contrast, failed calls to lightweight inspection commands such as \texttt{python}, \texttt{find}, and \texttt{grep} terminate early and are substantially cheaper. The resource cost of a failed call is therefore not uniformly low. Frameworks whose recovery paths repeatedly invoke \texttt{pip} or \texttt{pytest} accumulate non-trivial overhead per failed invocation regardless of whether those calls succeed.

\subsection{Resource Dynamics }
\label{Sec:Resource Decoupling}

\subsubsection{\textbf{Resource Dimensions Are Decoupled}}

The structural diversity observed in tool realization suggests that agent resource pressure is inherently multi-dimensional. Because frameworks materialize identical tasks through divergent execution paths, host-side costs do not scale linearly across resource dimensions. We therefore examine whether any single run-level metric can serve as a reliable proxy for others. Figure~\ref{fig:local-metric-pairs} compares CPU time, peak RSS, and subprocess count directly, and every pair shows only a moderate or weak monotonic relationship. The Spearman correlation is 0.54 for CPU time versus peak RSS, 0.53 for CPU time versus subprocess count, and only 0.13 for peak RSS versus subprocess count. These relationships are positive but not tight enough for any one metric to act as a reliable surrogate for the others.

The scatter plots make the decoupling concrete. SWE-agent reaches the highest subprocess count by a wide margin but does not occupy a correspondingly high position in peak RSS, illustrating why the RSS-subprocess correlation falls near zero. MetaGPT combines high CPU time with both elevated subprocess count and high peak RSS, placing it as an outlier across all three dimensions rather than along a single axis. Most other frameworks cluster at the lower end of all three metrics, but even within that cluster the relative ordering shifts depending on which dimension is measured. The ranking therefore changes by metric, which means CPU demand, memory footprint, and subprocess creation expose different aspects of host-side cost, and the rest of the paper treats these dimensions separately. 




\begin{figure*}[t!]
    \centering
    \includegraphics[width=0.99\linewidth]{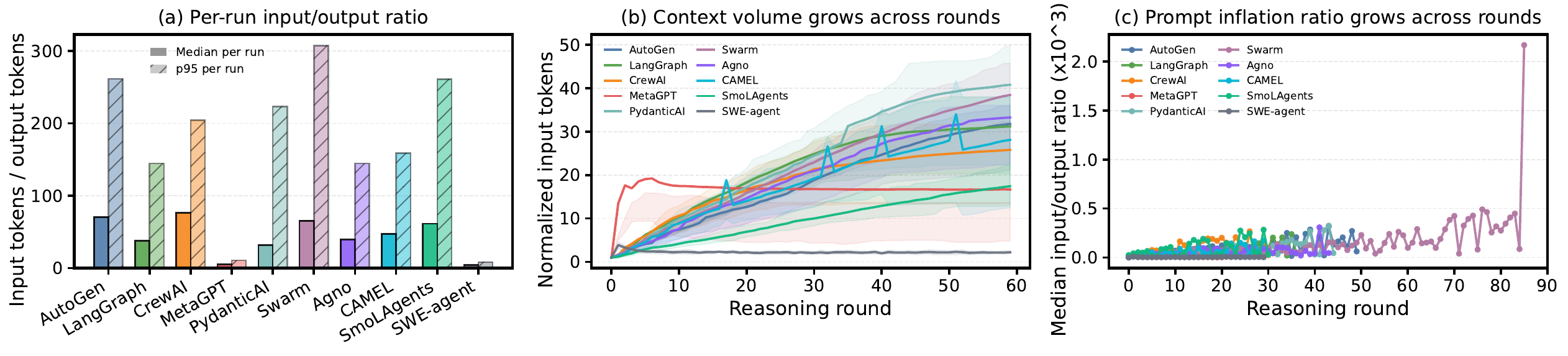}
    \Description{Three panels compare median and 95th-percentile input-to-output token ratios by framework, normalized input tokens across reasoning rounds, and input-to-output ratios across rounds.}
    \caption{Prompt-size and context-growth metrics across agent frameworks: (a) per-run input/output token ratio, including median and p95 values for each framework; (b) normalized input tokens across reasoning rounds; and (c) input/output token ratio across reasoning rounds.}
    \label{fig:tool-prompt-inflation}
\end{figure*}

\subsubsection{\textbf{Runtime State Accumulation Shapes Memory and Context Growth}}

Once the local metrics are separated, memory emerges as the most distinctive resource dimension. Figure~\ref{fig:memory-composition} shows that peak RSS is overwhelmingly anonymous rather than file-backed across all frameworks, with the anonymous share ranging from 84\% in Swarm to 97\% in CrewAI and file-backed memory remaining negligible in all cases. This does not prove that every byte represents semantically meaningful framework state, but it does establish that peak memory pressure is not a benign file-cache artifact from repository traversal. The dominant component is heap-like runtime state.


This pattern is consistent with how agent frameworks are implemented. Agent runtimes keep conversation history, tool outputs, intermediate plans, structured controller state, and wrapper-level objects alive across rounds, while also staging prompt assembly and response parsing in user-space data structures. Frameworks that retain more intermediate state or build thicker wrapper-side control paths therefore accumulate larger anonymous footprints.

\begin{figure}[t!]
    \centering
    \includegraphics[width=0.99\linewidth]{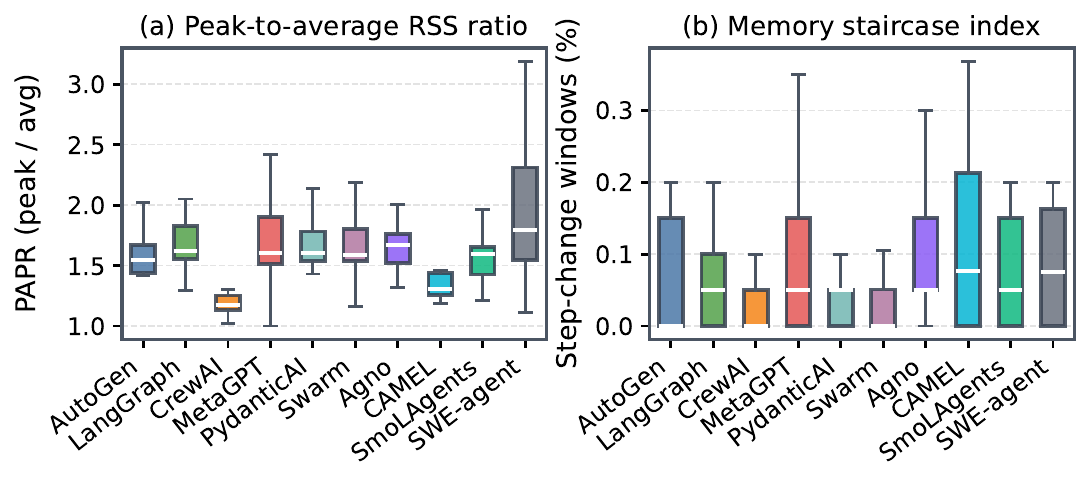}
    \Description{Two panels compare distributions of peak-to-average RSS ratios and memory staircase indices across ten frameworks, distinguishing transient bursts from stepwise accumulation.}
    \caption{Memory pressure differs in both transient burstiness and staircase-like accumulation across frameworks.}
    \label{fig:memory_lifecycle}
    \vskip -15pt
\end{figure}

The same state-retention behavior also changes what later model requests look like. Frameworks that preserve more prior context tend to re-inject more of that context into subsequent prompts, so local state growth and remote prompt growth often rise together. Figure~\ref{fig:tool-prompt-inflation}(a) shows substantial variation in per-run input/output token ratios, with Agno, SmoLAgents, and LangGraph reaching the highest tail ratios and CAMEL sustaining a consistently high median ratio. Figure~\ref{fig:tool-prompt-inflation}(b) shows that context volume grows monotonically across reasoning rounds, with CAMEL exhibiting the steepest growth trajectory.
Figure~\ref{fig:tool-prompt-inflation}(c) shows that CAMEL's prompt inflation ratio rises sharply at later rounds, reaching values far above all other frameworks as accumulated history is re-injected into each subsequent request. This pattern is consistent with framework implementation choices. Frameworks that preserve longer histories, repeatedly summarize prior tool outputs back into the prompt, or coordinate through layered role messages naturally expand request payload over time. Prompt inflation is therefore not a separate remote phenomenon. It is another surface manifestation of state carryover inside the runtime.

Figure~\ref{fig:memory_lifecycle} adds a temporal view that peak RSS alone misses. Figure~\ref{fig:memory_lifecycle}(a) shows that SWE-agent has the highest peak-to-average RSS ratio with the widest distribution, indicating that its memory risk comes from infrequent but sharp expansions. MetaGPT also shows a wide PAPR spread. CAMEL and CrewAI have the lowest PAPR values, indicating less bursty memory behavior. Figure~\ref{fig:memory_lifecycle}(b) reveals the complementary pattern. CAMEL has the highest median staircase index, indicating that a larger share of its time windows contain step-like memory increases consistent with steady accumulation across rounds. SmoLAgents and AutoGen also show elevated staircase indices, while CrewAI, PydanticAI, and Swarm remain low on both dimensions. These two dimensions capture distinct memory risk profiles. SWE-agent's memory risk arises from occasional large spikes, whereas CAMEL's risk arises from gradual but persistent accumulation. Some runtimes aggressively materialize large working sets around tool use or planner transitions, producing spikes, whereas others retain histories and controller objects more steadily across rounds, producing staircase-like growth.



\subsubsection{\textbf{Framework Sensitivity Varies across Resource Dimensions}}

\begin{figure}[t!]
    \centering
    \includegraphics[width=0.99\linewidth]{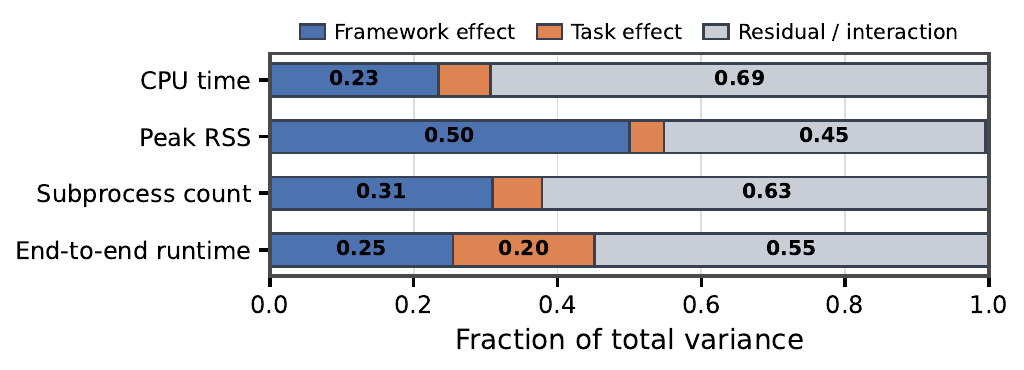}
    \Description{Horizontal stacked bars divide variance in CPU time, peak RSS, subprocess count, and end-to-end runtime into framework, task, and residual or interaction components.}
    \caption{Variance decomposition of host-side metrics into framework, task, and residual components.}
    \label{fig:framework-task-decomposition}
    \vskip -10pt
\end{figure}

We apply an additive variance decomposition over framework means and task means, with the remaining variance attributed to a residual term that reflects run-level execution variation. Figure~\ref{fig:framework-task-decomposition} shows the fraction of variance in each host-side metric explained by framework, task, and the residual term. Peak RSS shows the strongest framework effect at 0.50, with a small task component and a residual of 0.45. Subprocess count is also more strongly explained by framework than by task, at 0.31, with a larger residual of 0.63. CPU time has a smaller framework component of 0.23, with the residual dominating at 0.69. End-to-end runtime is the most distributed across all three terms, with framework at 0.25, task at 0.20, and the residual accounting for the remaining 0.55.  

This pattern reflects the degree to which each metric is governed by framework-level implementation decisions rather than by task content or run-level variation. Memory footprint and subprocess count are the most framework-determined metrics because both depend directly on how a framework retains state across rounds, structures its wrapper and controller paths, and materializes tool execution. CPU demand is less framework-determined because it reflects both the framework's realization path and the specific tool sequence a task triggers, mixing framework and task contributions into a large residual. End-to-end runtime shows the largest task component among all four metrics because task difficulty directly shapes the number of reasoning rounds required and therefore the total wall-clock time. Its residual is also the largest, driven primarily by API inference latency that varies per call and is determined neither by framework design nor by task content.

Together, these findings characterize agent workloads as both multi-dimensional and state-heavy. Host-side costs span weakly coupled resource dimensions, so no single metric serves as a reliable proxy for overall resource pressure. State accumulation drives the dominant components of memory and context growth in a framework-dependent manner, making framework choice the most actionable lever for managing resource demand. 

\section{Mismatch Analysis}
\label{Sec:Mismatch}

\subsection{Static Allocation Ignores Phase Structure}

Traditional resource management paradigms assume that a workload's resource demand is tightly coupled with a stable execution unit. For batch jobs\cite{dean2008mapreduce, zaharia2012resilient}, dominant computation spans the majority of the job lifecycle, making resource decisions at admission time effective throughout execution. For request-driven services\cite{barroso2003web} and serverless functions\cite{shahrad2020serverless, jonas2019cloud}, the request or invocation boundary serves the same role because execution paths are short and resource footprints are concentrated within that bounded interval. These abstractions hold when resource pressure either persists steadily across the workload lifetime or aligns cleanly with request boundaries. They further assume that CPU and memory scale together\cite{ghodsi2011dominant, grandl2016graphene, heo2015cgroupv2}, treating them as a jointly managed pair under the premise that more active work means proportionally more demand on both dimensions simultaneously.


However agents satisfy neither assumption. On the temporal axis our analysis shows that a single run repeatedly alternates between API wait intervals and local execution bursts. During remote model inference host side computation is suspended and activity drops sharply but once tool execution begins resource pressure spikes abruptly and recedes within seconds. Unlike DAG systems where phase structure is predefined agent phases are dynamically determined by model outputs making the timing of each burst unpredictable. On the resource axis CPU and memory decouple structurally across these phases. Memory accumulates across rounds and must remain resident during API wait phases to preserve conversation history and framework state even as CPU demand drops to zero. As shown in Figure~\ref{fig:local-metric-pairs} the Spearman correlation between CPU time and peak RSS is only 0.54 confirming that CPU and memory are not a jointly scalable resource pair.

This structural decoupling exposes the architectural rigidity of conventional allocation schemes. Because current controllers lack a representation of execution phases they cannot independently release CPU headroom during api wait without compromising the memory resident state required for subsequent rounds. This forces a false dichotomy under co location where provisioning for peak demand leads to massive resource under utilization while average based provisioning triggers contention and tool failure during unpredictable bursts. The mismatch is not a matter of numerical tuning but a fundamental lack of a control model that acknowledges the asymmetric retention semantics of agent phases. In this environment memory requires continuous protection while CPU permits aggressive and intermittent reclamation.

\subsection{Controllers Unaware of Resource Patterns}
Effective resource provisioning requires knowing not just how much a workload consumes, but how that consumption is generated.  For traditional workloads this distinction rarely matters because the execution path is stable and predictable. A database service typically materializes as heap-dominated memory growth, and a batch job always materializes as sustained core occupancy. Knowing the consumption amount is sufficient because the realization path is fixed.                                                   

For heterogeneous agent frameworks, the realization path varies fundamentally across frameworks even when the high-level agent semantics are identical. One framework may materialize tool execution through in-process calls while another spawns and tears down subprocesses repeatedly. MetaGPT triggers 117.2 execve events per tool call compared with only 3.0 in CAMEL, a 39x gap for the same high-level tool invocation. One framework may retain state compactly while another accumulates memory monotonically across rounds, with peak RSS variance explained 52.95\% by framework identity versus only 4.22\% by task content. These structural differences directly determine which resource dimensions will face pressure and when, but they are properties of the framework itself, not of any particular run.

A controller that lacks prior knowledge of these structural properties cannot make correct provisioning decisions. It cannot anticipate that one framework will require elevated subprocess capacity during tool execution while another requires protected memory headroom throughout API-wait phases. More fundamentally, without a framework fingerprint, the controller cannot interpret the origin of an observed pressure signal. A memory spike in MetaGPT most likely reflects its baseline orchestration footprint and warrants headroom protection; the same signal from a lighter runtime is more likely a transient retry artifact and warrants a different response. This ambiguity is compounded under co-location, where a single host runs multiple heterogeneous frameworks simultaneously and contention signals from structurally different sources are conflated into one shared resource view. Without per-framework structural knowledge, the controller has no basis for differentiated response and must fall back to reactive, uniform adjustments\cite{grandl2016graphene} triggered after degradation has already occurred, rather than pre-positioning resources according to the behavioral fingerprint of each co-located framework.

\subsection{Offline Profiling Misses Runtime Structure}

Our variance decomposition shows that run-level resource consumption in agent workloads is shaped by three sources operating at fundamentally different timescales. Framework structural properties such as memory growth model and subprocess realization path are relatively stable across tasks and runs, contributing a consistent structural fingerprint. Task content drives the concrete tool sequence triggered during execution and introduces cross-run variance that neither framework nor task identity can absorb; end-to-end runtime alone leaves 54.31\% of its variance unexplained by either factor. Within-run retry dynamics add a third dimension, causing memory to accumulate progressively within a single execution rather than converging to a stable working set. Because these three sources respond to different interventions and operate at different granularities, a single historical consumption figure cannot faithfully represent any of them in isolation.                                                                                                                
                              
Profiling-based allocation systems such as Borg~\cite{verma2015borg} and Autopilot~\cite{rzadca2020autopilot} derive resource recommendations from historical utilization percentiles, an approach that is well-grounded when the dominant source of consumption variance is a single stable factor such as request rate or input size. Under that condition, a P95 figure is interpretable because it reflects a genuine upper tail of the workload characteristic and can be expected to generalize to future runs. Agent workloads undermine this condition in a compounding way. When three entangled sources are folded into a single consumption figure, the resulting value is not a stable workload characteristic but an artifact of which factor happened to dominate a particular run in the profiling sample. A high P95 reading may reflect a framework with structurally heavy memory growth, a run that triggered an expensive tool sequence, or a retry spiral that inflated memory mid-run; nothing in the figure
distinguishes these cases. This ambiguity is not merely cosmetic, because the three cases call for qualitatively different allocation responses. Framework-structural cost benefits from pre-provisioning that is set once per framework deployment. Task-driven variance requires per-task adaptation that no single aggregate figure can capture. Retry-induced accumulation requires within-run elasticity that static percentiles cannot provide. A system that derives allocations from historical percentiles without disentangling these sources will over-provision for frameworks whose high P95 is driven by rare retry events and under-provision for frameworks whose structural memory growth is masked by low-cost tasks in the profiling window. Historical profiling can rank frameworks by average magnitude but cannot serve as a sound basis for differentiated allocation under a shared budget.

\section{Framework-Aware Resource Control} 
The three mismatches in §5 share a common cause: existing controls assume agent workloads are temporally stable and framework-agnostic. AgentLens bridges this gap by unifying offline framework fingerprints with online phase-aware control, enforcing differentiated memory coordination through cgroup v2 without instrumenting framework internals.


\subsection{Framework Fingerprint} 

To address the absence of framework structural priors, Agent-
Lens derives a per-framework fingerprint from historical execution traces before the controller runs. The fingerprint encodes two complementary forms of information. The first captures categorical attributes derived from implementation structure, including whether tool execution is in-process or subprocess-heavy, whether memory growth is bounded or monotonic, and how strongly CPU and PID activity are coupled. These attributes identify which resource dimensions will face structural pressure regardless of task content. The second captures shape statistics that quantify how a framework's execution style manifests in practice, covering subprocess burstiness, memory variability, and failure sensitivity. Rather than defining absolute resource thresholds, these statistics govern the intensity and timing of adjustments made at runtime. MetaGPT, for example, receives a subprocess-heavy classification with high \texttt{pids\_burstiness} and full-history state retention, while CAMEL is classified with bound-
ed memory growth and substantially lower process intensity, a distinction that maps directly to the 39$\times$\,\texttt{execve} gap documented in \S\ref{Sec:Tool Execution}. Fingerprints are loaded once at controller startup and remain stable across runs.

\subsection{Resource Controller} 

To address the three mismatches identified in \S\ref{Sec:Mismatch}, the Resource Controller operates a 100\,ms control loop that consumes three inputs. The first is the offline framework fingerprint loaded at startup. The second is a real-time execution phase signal derived from host-visible events without instrumenting framework internals. A run enters \texttt{api\_wait} after a model request has been issued and before the response returns, \texttt{tool\_executing} once a tool-call interval begins, \texttt{retry\_spiral} when failures repeat in a pattern indicating recovery behavior rather than isolated error, and \texttt{planning} otherwise. This four-state schema is coarse by design, aligning with resource behavior boundaries rather than semantic task structure and requiring only API-boundary and tool-call markers already present in the instrumentation pipeline. The third input is PSI and eBPF-derived pressure telemetry that reflects live resource contention across co-located instances. Framework choice explains 52.95\% of peak RSS variance but only 24.51\% of CPU core-seconds variance (\S\ref{Sec:Resource Decoupling}), and our characterization further shows that RSS accumulates as long-lived state while CPU activity is transient and bursty. Memory rather than CPU is therefore the binding constraint for multi-tenant concurrency density, and the Resource Controller focuses on memory coordination. The current phase determines the direction of adjustment and the framework fingerprint determines its magnitude. Instances in \texttt{tool\_executing} receive stronger memory protection while instances in \texttt{api\_wait} release headroom to peers. Frameworks with monotonic memory growth retain higher \texttt{memory.low} floors across all phases. This per-phase reweighting produces a three-tier protection structure, \emph{protection}, \emph{baseline}, and \emph{sacrifice}, which we refer to as the tiered policy in the evaluation. Enforcement uses \texttt{memory.high} to trigger gradual reclaim rather than \texttt{memory.max}, preserving accumulated framework state and avoiding OOM kills. To prevent oscillation from rapid phase transitions, the controller applies hold timers that stabilize memory assignments across short-lived phase changes and enforces a cooldown period before reversing adjustment direction. 

\subsection{Evaluation} 

\begin{figure}[t!] 
\centering 
\includegraphics[width=1.0\linewidth]{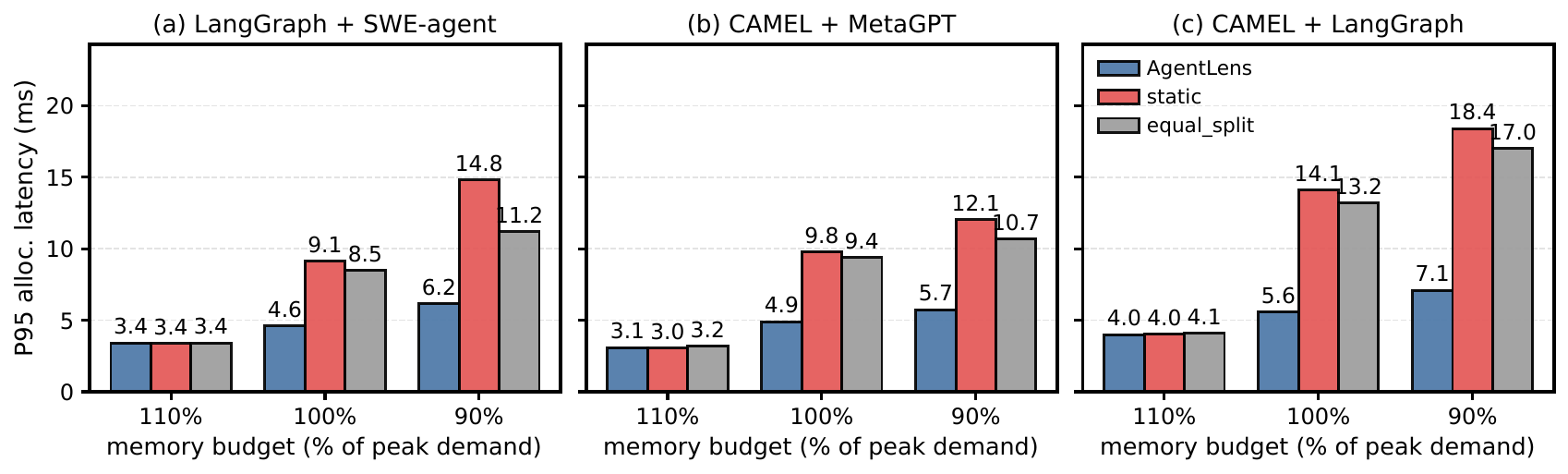} 
    \Description{Grouped bars compare AgentLens, static allocation, and equal split for three framework pairs at memory budgets of 110, 100, and 90 percent of peak demand. AgentLens has the lowest protected-worker P95 allocation latency at the tightest budget.}
\caption{Median P95 allocation latency of the protected worker under three coordinator modes, for three co-location pairs.} 
\label{fig:replay-eval-p95} 
\vskip -10pt
\end{figure} 

We evaluate AgentLens via trace-driven replay. Each worker replays a captured framework trace in its own cgroup~v2 child, allocates and releases anonymous memory along the recorded phase boundaries, and shares a parent slice whose \texttt{memory.max} is sized as a target fraction of the pair's aggregate peak demand. We compare three modes on identical traces and an identical hierarchy. \textsc{AgentLens} applies our phase-aware tiered policy, \textsc{static} holds a fixed per-worker budget, and \textsc{equal\_split} divides the parent budget evenly, so that any observed difference is attributable to the per-step \texttt{memory.low}/\texttt{memory.high} decisions alone. Three pairs cover distinct co-location regimes. LangGraph with SWE-agent (joint memory residency and subprocess churn), CAMEL with MetaGPT (light orchestration colocated with a process-rich multi-agent peer), and CAMEL with LangGraph in its context-memory variant (long-lived planning state on the peer). Each cell is repeated three times at parent budgets equal to 110\%, 100\%, and 90\% of the pair's peak demand; 110\% bounds controller overhead under a loose budget, while 90\% places aggregate demand above the budget so that reclaim is continuously active. Figure~\ref{fig:replay-eval-p95} reports the median P95 allocation latency of the protected worker. At 110\%, the three modes agree within 6.7\% (3.0--4.1\,ms), confirming that AgentLens adds no measurable overhead when the parent budget is loose. At 90\%, \textsc{AgentLens} keeps the protected worker's P95 latency at 5.7--7.1\,ms across the three pairs, while \textsc{static} inflates to 12.1--18.4\,ms and \textsc{equal\_split} to 10.7--17.0\,ms, a $2.1\times$--$2.6\times$ and $1.8\times$--$2.4\times$ regression respectively. Two supporting observations follow from the per-step coordinator log: \textsc{static} fails to complete two to three of every three repetitions at 90\% because OOM-kill cascades through the context-heavy worker, whereas \textsc{AgentLens} completes all repetitions; and at least 90\% of \texttt{tool\_executing} samples are placed in the protection tier, confirming that the latency gains arise where the policy was designed to act.

\section{Conclusion}
\label{Sec:Discussion and Conclusion}
Our cross-framework characterization reveals that host-side resource behavior of LLM-based coding agents varies substantially across framework identities, even when task content is held comparable. Framework-level differences account for an order-of-magnitude greater peak memory variance than task-level differences, suggesting that orchestration structure is a non-negligible factor in resource dynamics. These findings motivate framework-aware resource abstractions. We demonstrate this principle through AgentLens, which exposes orchestration-induced resource patterns to enable targeted coordination. Limitations remain in generalizing across closed-source frameworks and fully automated fingerprinting. Addressing these requires deeper integration between agent runtime design and systems-level resource management.

\bibliographystyle{ACM-Reference-Format}
\bibliography{software}
\end{document}